\documentstyle[12pt,epsf]{article}

\def\thefootnote{\fnsymbol{footnote}}

\begin{document}

\newcommand{\be}{\begin{equation}}	
\newcommand{\ee}{\end{equation}}	
\newcommand{\beq}{\begin{eqnarray}}	
\newcommand{\eeq}{\end{eqnarray}}	
\newcommand{\beqstar}{\begin{eqnarray*}}	
\newcommand{\eeqstar}{\end{eqnarray*}}	

\newcommand{\gsim}{ \mathop{}_{\textstyle \sim}^{\textstyle >} }
\newcommand{\lsim}{ \mathop{}_{\textstyle \sim}^{\textstyle <} }
\newcommand{\vev}[1]{ \left\langle {#1} \right\rangle }
\newcommand{\lsp}{ \left ( }
\newcommand{\rsp}{ \right ) }
\newcommand{\lmp}{ \left \{ }
\newcommand{\rmp}{ \right \} }
\newcommand{\llp}{ \left [ }
\newcommand{\rlp}{ \right ] }
\newcommand{\labs}{ \left | }
\newcommand{\rabs}{ \right | }

\newcommand{\tr}{{\rm tr}} 
\newcommand{\K}{ {\rm K} }
\newcommand{\EV}{ {\rm eV} }
\newcommand{\KEV}{ {\rm keV} }
\newcommand{\MEV}{ {\rm MeV} }
\newcommand{\GEV}{ {\rm GeV} }
\newcommand{\TEV}{ {\rm TeV} }
\def\gtap{\raisebox{-.4ex}{\rlap{$\sim$}} \raisebox{.4ex}{$>$}}
\newcommand{\gggh}{${\rm G_{GUT}\times G_H}$ }

\begin{titlepage}
\begin{center}

\hfill    FERMILAB-PUB-97/019-T\\
\hfill    hep-ph/9702214 \\
\hfill    January, 1997

\vskip .5in

{ \Large \bf Supersymmetric Dynamical Generation\\
 of the Grand Unification Scale
}

\vskip .5in

{\large Hsin-Chia Cheng}\footnote{Email: hcheng@fnal.gov}

\vskip .5in

{\it Fermi National Accelerator Laboratory\\
     P.O. Box 500\\
     Batavia, IL 60510}

\end{center}

\vskip .5in

\begin{abstract}

The grand unification scale $M_{\rm GUT}\sim 10^{16}$GeV
may arise from dynamical effects. With the advances in 
understanding of supersymmetric dynamics, we can break
the grand unified group by introducing a strong gauge 
group which generates the grand unification scale. 
We also show how this mechanism can be combined with
solutions to the doublet-triplet splitting problem. The 
same method can also be used for other symmetry breakings 
at intermediate scales as well. 

\end{abstract}
\end{titlepage}

\renewcommand{\thepage}{\arabic{page}}
\setcounter{page}{1}
\renewcommand{\thefootnote}{\arabic{footnote}}
\setcounter{footnote}{0}

One of the outstanding questions in particle physics is the
``hierarchy problem'': why is the mass scale of ordinary particle 
physics so small compared to the gravitational scale? In 
supersymmetric (SUSY) theories, the electroweak scale is related
to the supersymmetry breaking scale, the problem is then translated
into why supersymmetry is broken at a scale much below the Planck
mass $M_{\rm pl}$. As Witten pointed out\cite{Witten}, if 
supersymmetry is dynamically broken by non-perturbative effects,
a large hierarchy between the SUSY-breaking scale and the Planck
scale can be naturally generated. Recently, Seiberg and his
collaborators have made great progress
in understanding non-perturbative SUSY
dynamics\cite{Seiberg}. Many more models of dynamical supersymmetry
breaking have been found, providing new hope to understand
how supersymmetry is broken in nature and how the electroweak scale
is generated.

In many extensions of the standard model (SM), there are additional
intermediate scales between the electroweak scale and the Planck
scale. One example is the grand unification scale. The fact that the
gauge couplings of SU(3)$_{\rm C}$, SU(2)$_{\rm W}$ and U(1)$_{
\rm Y}$ of the minimal SUSY extension of the standard model meet
together at about $10^{16}$GeV\cite{gc_unification} 
gives a non-trivial indication that a
SUSY GUT\cite{SUSY_GUT} may be realized in nature and the GUT gauge
group is
broken at $\sim 10^{16}$GeV, two orders of magnitude beneath
$M_{\rm pl}$.\footnote{The two scales are so close that there may
be ways to reconcile the discrepancy, e.g. in strong coupled
string theories\cite{Witten2}.}
In flavor theories which try to understand the
fermion mass hierarchy, the Froggatt-Nielsen mechanism\cite{FN}
is often used to generate small numbers from the ratios of 
different mass scales. This also requires flavor symmetry
breaking at intermediate scales. Similar to the SUSY-breaking
scale, these intermediate symmetry breaking scales may also arise 
from dynamical effects. We will try to use the advances in
our understanding of SUSY dynamics to generate these symmetry
breaking scales dynamically. 
Our philosophy is that no explicit
mass parameter (except maybe $M_{\rm pl}$) should appear in the 
fundamental Lagrangian.
There are other ways of generating intermediate scales, for
example, soft SUSY-breaking scalar mass squareds can be driven
negative by Yukawa couplings\cite{HG}. However, understanding the origin
of symmetry
breaking is one of the deepest problems we face, and it is important
to explore various possibilities.

Taking a simple example, consider the superpotential
\be
W=\lambda X (\bar{\phi} \phi -\mu^2), \label{sb}
\ee
where $\bar{\phi}, \phi$ transform under some symmetry group,
$X$ is a singlet, and $\mu$ is some mass parameter. The equation
$\partial W/ \partial X =0$ will force  $\bar{\phi}, \phi$ to 
get vacuum expectation values (VEVs) 
and break the symmetry at the scale $\mu$. We would like
to have the scale $\mu$ generated dynamically instead of being
put in by hand. The simplest thing one may try to do is to
replace $\mu^2$ by $\bar{Q} Q$, where $\bar{Q}, Q$ transform
under some strong gauge group and form a condensate. However,
the extra coupling $X \bar{Q} Q$ often disturbs the original
vacuum and may generate a runaway direction in which $\vev{X}
\to \infty$ and $\langle \bar{Q}Q\rangle \to 0$. Adding $X^3$
interaction removes the runaway direction but also forces
$\langle \phi \rangle = \langle \bar{\phi} \rangle =0$.
One can cure 
this by introducing an additional singlet. Consider a strong
gauge group SU(N) with one flavor $\bar{Q}$, $Q$ and a singlet
$S$. The tree level superpotential is given by
\be
W_{\rm tree}=\lambda_1 S\bar{Q}Q + {\lambda_2 \over 3} S^3.
\ee
Because $N>1$, a nonperturbative superpotential is generated
dynamically\cite{ADS},
\be
W_{\rm dyn}= (N-1) \left({\Lambda^{3N-1} \over 
\bar{Q} Q}\right)^{1 \over N-1}.
\ee
Integrating out  $\bar{Q}, Q$, (they get mass from the VEV of S, which
is justified below,)
the first term in $W_{\rm tree}$
together with $W_{\rm dyn}$ will generate a runaway superpotential for
the singlet S,
\be
N(\lambda_1 S \Lambda^{3N-1})^{1\over N}
\ee
It will be stablized by the second term in $W_{\rm tree}$.
Solving the equation of motion we find that $S$ gets a VEV of
the order of the strong SU(N) scale,
\be
\vev{S}= \Lambda \left(-{\lambda_1 \over \lambda_2^N}\right)^{1\over 3N-1}.
\ee
Now we can replace $\mu^2$ in (\ref{sb}) by $S^2$, and the coupling
$X S^2$ will not destroy the original vacuum. In this way, we
can break symmetry groups dynamically and generate the symmetry
breaking scale from strong dynamics. 

The superpotential in the above discussion is not the most
general one which can be written down. The most general superpotential
without any mass parameter is\footnote{Planck scale mass terms
are assumed to be absent or may be forbidden by some discrete symmetry.
Non-renormalizable terms suppressed by $M_{\rm pl}$ are smaller
and should have little effect on the vacua.}
\be
W=\lambda_1 S \bar{Q} Q+ {\lambda_2 \over 3} S^3 
+{\lambda_3 \over 2} S^2 X + {\lambda_4 \over 2} S X^2
+ {\lambda_5 \over 3} X^3 + (\lambda_6 S +\lambda_7 X)
\bar{\phi} \phi. \label{dsb}
\ee
(The $X \bar{Q} Q$ coupling can be removed by redefining
$X$ and $S$.) Including the nonperturbative superpotential
and solving the equations of motion, one finds that vacua
with $\langle \bar{\phi} \rangle, \vev{\phi}=0$ and
$\langle \bar{\phi} \rangle, \vev{\phi} \neq 0$ both
exist. Multiple vacua are a generic feature of the
supersymmetric theories.
If we sit on the $\langle \bar{\phi} \rangle, \vev{\phi} \neq 0$
vacuum, the symmetry is broken dynamically.

In the rest of this letter, we will concentrate on
breaking the GUT symmetry dynamically. The apparent unification
of the standard model gauge couplings at $M_{\rm GUT} \approx
10^{16} \GEV$ makes it the most possible intermediate scale to 
exist between $M_{\rm pl}$ and $M_W$. However, the method
discussed here can be applied to other symmetry breakings as well.

Consider a GUT model based on the gauge group 
SU(N)$\times$SU(5)$_{\rm GUT}$ with $N>5$, where SU(5)$_{\rm GUT}$
is the ordinary grand unified group, and SU(N) is strongly coupled
with scale $\Lambda \sim M_{\rm GUT}$. To break SU(5)$_{\rm GUT}$,
the model contains the fields $Q^i_\alpha$, $\bar{Q}_i^\alpha$,
which transform like (${\bf N, 5^*}$) and (${\bf N^*, 5}$) under
SU(N)$\times$SU(5)$_{\rm GUT}$, and $\Sigma_i^j$, which transform
like ($\bf{1, 24}$), where $\alpha=1, \cdots, N$ and $i, j=1, \cdots, 5$
are SU(N) and SU(5)$_{\rm GUT}$ indices respectively.
The tree level superpotential is given by
\be
W_{\rm tree}= \lambda_1 Q \Sigma \bar{Q} + {\lambda_2 \over 3}
{\rm tr}(\Sigma^3).
\ee
These are the only terms one can write down without any mass
parameter. For the SU(N) gauge group, the number of flavors
$N_f=5$ is less than the number of colors, so a nonperturbative
superpotential is dynamically generated\cite{ADS},
\be
W_{\rm dyn}= (N-5) \left({\Lambda^{3N-5} \over \det Q\bar{Q}}\right)^{
1\over N-5}, \label{npert}
\ee
where the determinant is taken on the flavor (SU(5)$_{\rm GUT}$) indices.
Combining it with the tree-level superpotential we can look
for the vacua. There is one runaway vacuum which preserve
SU(5)\footnote{In the Higgs picture, the unbroken SU(5) may be thought
as a combination of SU(5)$_{\rm GUT}$ and a subgroup of SU(N)}:
$\vev{\Sigma}=0,\, \langle \bar{Q}Q\rangle \propto {\bf 1}
\to \infty$. We are interested in vacua in which $\vev{\Sigma}
\neq 0$ and SU(5) is broken. To study them we can integrate
out $\bar{Q},Q$ and obtain an effective superpotential depending
only on $\Sigma$,
\be
W_{\rm eff}= N \Lambda^{3N-5\over N} (\lambda_1^5 
\det \Sigma)^{1\over N} + {\lambda_2 \over 3}\tr \Sigma^3.
\ee
Solving $\partial W_{\rm eff}/ \partial \Sigma^i_j=0$,
we find only two inequivalent solutions,
\beq
&\vev{\Sigma}={1\over \sqrt{60}}
\left(
\begin{array}{ccccc}
2&0&0&0&0 \\ 0&2&0&0&0 \\ 0&0&2&0&0 \\ 0&0&0&-3&0 \\
0&0&0&0&-3
\end{array}
\right) v, \,\,\, 
v={\Lambda \over \lambda_1} ({10\sqrt{60} \lambda_1^3
\over \lambda_2})^{N\over 3N-5}({1\over 50 \sqrt{60}})^{
1\over 3N-5}, \\
{\rm and}\,
&\vev{\Sigma}={1\over \sqrt{40}}
\left(
\begin{array}{ccccc}
1&0&0&0&0 \\ 0&1&0&0&0 \\ 0&0&1&0&0 \\ 0&0&0&1&0 \\
0&0&0&0&-4
\end{array}
\right) u, \,\,\, 
u={\Lambda \over \lambda_1} (-{10\sqrt{40} \lambda_1^3
\over 3\lambda_2})^{N\over 3N-5}({1\over 400 \sqrt{40}})^{
1\over 3N-5}.
\eeq
In these vacua, SU(5)$_{\rm GUT}$ is dynamically broken
down to SU(3)$\times$SU(2)$\times$U(1) and SU(4)$\times$U(1)
respectively. The GUT scale is generated by the strong
SU(N) group. If all Yukawa couplings are $\cal O$(1), then
we have $M_{\rm GUT} \sim {\cal O}(\Lambda)$.\footnote{
A worry is that when incorporated into supergravity,
the strong gauge dynamics may break supersymmetry\cite{Nilles},
then the supersymmetry breaking scale will be too high. However,
how to correctly incorporate supergravity is a complicated issue
and worth further investigation.
Without knowing it exactly, we
will assume that supersymmetry is broken by some other sector,
not by this strong SU(N) gauge group.}

If the fields get some soft SUSY-breaking masses after 
supersymmetry is broken, the degeneracy among these vacua
will be lifted. One expect that the runaway vacuum will be 
disfavored if the soft SUSY-breaking mass squares are positive.
In the runaway direction, the supersymmetric contribution to
the potential scales as 
\be
V_{\rm SUSY} \sim \left({\Lambda^{3N-5}\over v^{10}}\right)^{2\over N-5}
{1\over v^2},
\ee
where $v$ is the VEV of $Q$ and $\bar{Q}$. It is stablized by
the soft breaking terms,
\be
V_{\rm soft} \sim m_s^2 v^2,
\ee
where $m_s$ is the soft breaking mass. The minimum occurs
when these two terms are balanced,
\be
 \left({\Lambda^{3N-5}\over v^{10}}\right)^{2\over N-5} {1\over v^2}
\sim m_s^2 v^2.
\ee
Then, the VEVs of $Q$ and $\bar{Q}$ scale as $v\sim \lambda
(\Lambda/m_s)^{N-5\over 2N}$
and the vacuum energy scale as $m_s^{N+5\over N}
\Lambda^{3N-5\over N}$.\footnote{For 
some $N$ and $m_s$, $v$ can be larger than 
$M_{\rm pl}$, and then one may not trust this result. However,
we just use these results to get a rough idea of the energy of 
the runaway vacuum and we will ignore this problem.}
Compared with the SU(5) breaking minima, which have energy
$V\sim m_s^2 \Lambda^2$, the runaway vacuum clearly has higher
energy for $N>5$ (and $\Lambda > m_s$). The K\"{a}hler potential
is unknown so we can not compare the vacuum energies at the
SU(3)$\times$SU(2)$\times$U(1) and SU(4)$\times$U(1) minima.
We assume that the SU(3)$\times$SU(2)$\times$U(1) minimum
is somehow chosen by nature.

We can modify this model by adding a singlet $S$,
coupled to $\bar{Q}, Q$, to lift the runaway direction. The 
most general tree-level superpotential without any mass
parameter is
\be
W_{\rm tree}= \lambda_1 Q \Sigma \bar{Q} + {\lambda_2 \over 3}
\tr \Sigma^3 + \lambda_3 S Q \bar{Q} + {\lambda_4 \over 3}S^3
+{\lambda_5 \over 2}S\, \tr \Sigma^2. \label{ws}
\ee
Again, with the nonperturbative superpotential (\ref{npert}),
there are several discrete vacua, among them the
 desirable one breaking SU(5) down to
SU(3)$\times$SU(2)$\times$U(1). There are other vacua which
preserve SU(5) or break it down to SU(4)$\times$U(1),
SU(3)$\times$U(1)$^2$, and SU(2)$^2 \times$U(1).
Both $S$ and $\Sigma$ get VEVs of the order $\Lambda$
except in the SU(5) preserving vacuum in which $\vev{\Sigma}
=0$.

One of the most serious problem of the grand unified theories
is the ``doublet-triplet splitting problem''. Higgs doublets are
responsible for the electroweak symmetry breaking and hence 
their masses are of the order of the weak scale. On the other
hand, their color-triplet partners must have GUT scale masses
in order to achieve successful gauge coupling unification
and/or to avoid rapid proton decay. In the  minimal
SUSY SU(5), this hierarchy is obtained by an extreme fine
tune of the parameters in the superpotential, which is
obviously unsatisfactory. In the following we will
combine the dynamical GUT breaking model with some known
solutions to the doublet-triplet splitting problem.

The first solution we consider is the pseudo-Goldstone boson
mechanism\cite{PGB}. It is based on the gauge group SU(6).
The SU(6) is broken down to the standard model gauge group
by two kinds of Higgs representations, an adjoint $\Sigma$
with the VEV,
\be
\vev{\Sigma}={\rm diag} (1, 1, 1, 1, -2, -2) v,
\ee
and a fundamental-antifundamental pair $H$ and $\bar{H}$
with the VEVs
\be
\vev{H}=\langle\bar{H}\rangle =(a, 0, 0, 0, 0, 0).
\ee
If there is no cross coupling between $\Sigma$ and $H, \bar{H}$
in the superpotential,
\be
W=W(\Sigma)+W(H, \bar{H}),
\ee
then there is an effective SU(6)$_{\Sigma} \times$SU(6)$_H$
symmetry. These two SU(6)'s are broken down to 
SU(4)$\times$SU(2)$\times$U(1) and SU(5) respectively.
By a simple counting of the Goldstone modes and the 
broken gauge generators, one can find there are two electroweak
doublets not eaten by the gauge boson and hence left massless.
They are linear combinations of the $\Sigma$ and $H, \bar{H}$
fields,
\be
{a H_{\Sigma} - 3v H_H \over \sqrt{a^2 +9v^2}}, \,\,\,
{a \bar{H}_{\Sigma} - 3v \bar{H}_{\bar{H}} \over \sqrt{a^2 +9v^2}},
\ee
and they will get weak scale masses from radiative corrections.

The simplest way to generate $v$ and $a$ dynamically is to use
the dynamical model with the singlet (\ref{ws}), replacing SU(5)
by SU(6). The field content contains $Q, \bar{Q}$, $S$, $\Sigma$,
$H, \bar{H}$ discussed above, and a additional singlet $X$.
The superpotential is given by
\be
W= \lambda_1 Q \Sigma \bar{Q} + {\lambda_2 \over 3}
\tr \Sigma^3 + \lambda_3 S Q \bar{Q} + {\lambda_4 \over 3}S^3
+{\lambda_5 \over 2}S\, \tr \Sigma^2 + \lambda_6 X \bar{H} H
- \lambda_7 X S^2.
\ee
Similar to what we discussed before, there is a vacuum in 
which both $\vev{S}, \vev{\Sigma} \sim {\cal O}(\Lambda)$
and $\vev{\Sigma}$ takes the form
\be
\vev{\Sigma}={\rm diag} (1, 1, 1, 1, -2, -2) v.
\ee
The last two terms in the superpotential will force $H, \bar{H}$
to get ${\cal O}(\Lambda)$ VEVs. Therefore, the pseudo-Goldstone
boson mechanism can be achieved with both $\Sigma$ and $H, \bar{H}$
VEVs generated dynamically, and their scales naturally tied together.

The problem with this model is that there is no explanation for
the absence of the $\bar{H}\Sigma H$ coupling. This coupling,
if it exists, destroys the pseudo-Goldstone boson mechanism.
Although it is technically natural to omit this coupling in
supersymmetric theories, one may prefer to having some symmetry
reason to forbid this coupling. There are no such symmetries
in this model. One possibility is to generate $H, \bar{H}$
VEVs from another sector. (Then we lose the natural link between
the two scales.)
For example, we can use the method
 discussed before: Introducing another gauge group
SU(M) with one flavor of fundamental and antifundamental fields
 and generating $H$ and $\bar{H}$ VEVs through the superpotential
(\ref{dsb}). In this case, we can assign separate $Z_3$ symmetries
to the two sectors which generate the $\Sigma$ and $H, \bar{H}$
VEVs. The lowest order nonrenormalizable coupling suppressed
by $M_{\rm pl}$ between $\Sigma$ and $H, \bar{H}$ allowed
will be
\be
{(\bar{H}\Sigma H)^3 \over M_{\rm pl}^6}.
\ee
The induced masses for the light Higgs doublets will be
$M_{\rm GUT}(M_{\rm GUT}/M_{\rm pl})^6$ and no bigger than
the weak scale if $M_{\rm GUT}/M_{\rm pl} \lsim 1/200$.
Another way is to use the anomalous U(1) symmetry to generate
$H, \bar{H}$ VEVs and forbid $\bar{H} \Sigma H$ coupling,
as discussed in \cite{DP}.

Another solution to the doublet-triplet splitting problem
which we consider is proposed by Yanagida
et al.\ \cite{Yana,Yana1,Yana2}.
Let us first review the idea using the model given in
\cite{Yana1}. The model is based on the gauge group
SU(5)$_{\rm GUT} \times$SU(3)$_{\rm H} \times$U(1)$_{\rm H}$, with
the following fields,
$R({\bf 5^*}, {\bf 3}, 1)$, $\bar{R}({\bf 5}, {\bf 3^*}, -1)$, $q({\bf
1}, {\bf 3}, 1)$, $\bar{q}({\bf 1}, {\bf 3^*}, -1)$, $\Sigma ({\bf
24}, {\bf 1}, 0)$, $H({\bf 5}, {\bf 1}, 0)$, and $\bar{H}({\bf 5^*},
{\bf 1}, 0)$, where the numbers in brackets denote the transformation
properties under SU(5)$_{\rm GUT}$, SU(3)$_{\rm H}$ and 
U(1)$_{\rm H}$, respectively. One can write down a superpotential,
\beq
W=m_R\, \tr( R\bar{R}) +\lambda R \Sigma \bar{R} + {1\over 2}m_{\Sigma}\,
\tr (\Sigma^2) +h H R \bar{q} +h' \bar{H} \bar{R} q,
\eeq
which has a vacuum as follows,
\beq
&&\langle R \rangle = \langle \bar{R}^{\rm T} \rangle = 
\left(
\begin{array}{ccccc}
1&0&0&0&0 \\ 0&1&0&0&0 \\ 0&0&1&0&0
\end{array}
\right) \sqrt{5m_R m_{\Sigma} \over \lambda^2},
 \nonumber \\
&&\langle\Sigma\rangle =
{m_R \over 2\lambda} {\rm diag}(2, 2, 2, -3, -3).
\label{yanavev}
\eeq
In this vacuum the gauge symmetry 
SU(5)$_{\rm GUT} \times$SU(3)$_{\rm H} \times$U(1)$_{\rm H}$ 
is broken down
to the standard model gauge group 
SU(3)$_{\rm C}\times$SU(2)$_{\rm W}\times$U(1)$_{\rm Y}$.
The terms $H R \bar{q}$ and $\bar{H}\bar{R} q$ in the superpotential
marry the color-triplet Higgses in $H, \bar{H}$
to $\bar{q}$ and $q$ so that they obtain the GUT scale masses
while the doublet Higgses remain massless. The doublet-triplet
splitting is achieved by the missing partner mechanism with 
small matter representations.

The low energy SU(3)$_{\rm C}$ and U(1)$_{\rm Y}$ are diagonal subgroups
of the SU(3), U(1) in SU(5)$_{\rm GUT}$ and the SU(3)$_{\rm H}$,
U(1)$_{\rm H}$ groups in this model. The gauge coupling unification
is not spoiled if the gauge couplings of the SU(3)$_{\rm H}$ and the
U(1)$_{\rm H}$ are big enough. In fact, the corrections from the
SU(3)$_{\rm H}$ and U(1)$_{\rm H}$ couplings lower the prediction for
the strong coupling constant
$\alpha_S$ in SUSY GUT and therefore move it in the right
direction\cite{ACM}.

To combine it with the dynamical GUT breaking model, we consider the gauge
group SU(N)$\times$SU(5)$_{\rm GUT} \times$SU(3)$_{\rm H}
\times$U(1)$_{\rm H}$, with the following field content:
$Q({\bf N, 5^*, 1}, 0)$, $\bar{Q}({\bf N^*, 5, 1}, 0)$,
$\Sigma ({\bf 1}, {\bf 24}, {\bf 1}, 0)$,
$R({\bf 1},{\bf 5^*}, {\bf 3}, 1)$, $\bar{R}({\bf 1},{\bf 5}, {\bf 3^*}, -1)$,
$q({\bf 1}, {\bf 1}, {\bf 3}, 1)$, $\bar{q}({\bf 1}, {\bf 1}, {\bf 3^*}, -1)$, 
$H({\bf 1},{\bf 5}, {\bf 1}, 0)$, $\bar{H}({\bf 1}, {\bf 5^*},
{\bf 1}, 0)$, and $S({\bf 1, 1, 1}, 0)$.
The tree-level superpotential is given by
\be
W=\lambda_1 Q \Sigma \bar{Q} + {\lambda_2 \over 3} \tr (\Sigma^3)
+ \lambda_3 R \Sigma \bar{R} - \lambda_4 S\, \tr (R \bar{R})
+{\lambda_5 \over 3} S^3 + h H R \bar{q} + h' \bar{H}\bar{R} q.
\ee
We are interested in the vacuum in which $\vev{\Sigma}, \vev{R},
\langle \bar{R} \rangle \neq 0$. Integrating out $Q, \bar{Q}$, we obtain
\be
W_{\rm eff}= N (\lambda_1^5 \det \Sigma)^{1\over N}
\Lambda^{3N-5\over N} +  {\lambda_2 \over 3} \tr (\Sigma^3)
+ \lambda_3 R \Sigma \bar{R} - \lambda_4 S\, \tr (R \bar{R})
+{\lambda_5 \over 3} S^3 + h H R \bar{q} + h' \bar{H}\bar{R} q.
\ee
Solving the equations of motion,
\beq
{\partial W_{\rm eff} \over \partial S}&=& -\lambda_4 \tr (R \bar{R})
+ \lambda_5 S^2 =0, \\
{\partial W_{\rm eff} \over \partial R}&=& \lambda_3 \Sigma \bar{R}
-\lambda_4 S \bar{R} =0, \\
{\partial W_{\rm eff} \over \partial \Sigma}&=& {\Lambda^{3N-5 \over N}
\over \lambda_1} (\lambda_1^5 \det \Sigma)^{1\over N}
(\Sigma^{-1}-{1\over 5}\tr (\Sigma^{-1}))  \nonumber \\
&&+ \lambda_2 (\Sigma^2-{1\over 5}\tr (\Sigma^2))
+\lambda_3 (R\bar{R} -{1\over 5}\tr (R\bar{R}))=0,
\eeq
one can find a vacuum with
\beq
&&\langle\Sigma\rangle =
{\rm diag}(2, 2, 2, -3, -3)  v, \nonumber \\
&&\langle R \rangle = \langle \bar{R}^{\rm T} \rangle = 
\left(
\begin{array}{ccccc}
1&0&0&0&0 \\ 0&1&0&0&0 \\ 0&0&1&0&0
\end{array}
\right) \times {2\lambda_3 \over \lambda_4} \sqrt{\lambda_5 \over 3 
\lambda_4} v, \nonumber \\
&& \vev{S}= {2\lambda_3 \over \lambda_4} v, 
\eeq
where 
\beq
v=72^{1\over 3N-5} \lambda_1^{5-N \over 3N-5} \left(6\lambda_2 -
{8 \lambda_3^3 \lambda_5 \over 5\lambda_4^3}\right)^{-{N \over 3N-5}}
\Lambda . \nonumber
\eeq
Thus, we obtain the desirable vacuum in the form of (\ref{yanavev})
and the missing partner mechanism for the doublet-triplet
splitting problem can be implemented.

In summary, we have shown that how the grand unified gauge group
can be dynamically broken down to the standard model gauge group 
without inputting the GUT scale explicitly
by hand. We also showed that this mechanism can be combined with solutions to 
the doublet-triplet splitting problem. Although we concentrated our
discussion on GUT symmetry breaking, the same method can also
be useful for other symmetry breakings at intermediate scales as 
well.

The author would like to thank L.~J.~Hall, J.~D.~Lykken, W.~Skiba,
Y.-Y.~Wu,
and especially N. Arkani-Hamed and S.~Trivedi for valuable
discussions and suggestions.
Fermilab is operated by Universities Research Association, Inc.,
under contract DE-AC02-76CH03000 with U.S. Department of Energy.


%
%
\newcommand{\Journal}[4]{{\sl #1} {\bf #2} {(#3)} {#4}}
\newcommand{\APJ}{Ap. J.}
\newcommand{\CJP}{Can. J. Phys.}
\newcommand{\NC}{Nuovo Cimento}
\newcommand{\NP}{Nucl. Phys.}
\newcommand{\MPL}{Mod. Phys. Lett.}
\newcommand{\PL}{Phys. Lett.}
\newcommand{\PR}{Phys. Rev.}
\newcommand{\PRep}{Phys. Rep.}
\newcommand{\PRL}{Phys. Rev. Lett.}
\newcommand{\PTP}{Prog. Theor. Phys.}
\newcommand{\SJNP}{Sov. J. Nucl. Phys.}
\newcommand{\ZP}{Z. Phys.}

\end{document}